\documentclass{aa}
\usepackage{graphicx}
\usepackage[longnamesfirst]{natbib}
\begin{document}
   \title{Millimeter-wave emission during the 2003 low excitation phase of $\eta$ Carinae}

   \author{Z. Abraham
          \inst{1}
          \and
          D. Falceta-Gon\c calves\inst{1}
		  \and
		  T. P. Dominici\inst{1}
		  \and
		  L.-\AA. Nyman\inst{2,3}
		  \and
		  P. Durouchoux\inst{4}
		  \and
		  F. McAuliffe\inst{2}
		  \and
		  A. Caproni\inst{1}
		  \and
		  V. Jatenco-Pereira\inst{1}
		  }

   \offprints{Z. Abraham}

   \institute{Instituto de Astronomia, Geof\'\i sica e Ci\^encias
   Atmosf\'ericas, Universidade de S\~ao Paulo, Rua do Mat\~ao 1226, Cidade
   Universit\'aria, 05508-900, S\~ao Paulo, Brazil\\
              \email{zulema@astro.iag.usp.br}
         \and
             European Southern Observatory, Casilla 19001, Santiago 19, Chile
		\and
		Onsala Space Observatory, 439 92 Onsala, Sweden
		\and
		Service d'Astrophysique (CNRS URA 2052), CEA Saclay, 9119Gif-sur-Yvette,
		France\\
			}

  \date{Received; Accepted }

   \abstract{
In this paper we present observations of $\eta$ Carinae in the 1.3 mm and 7 mm radio continuum, during 
the 2003.5 low excitation phase. 
The  expected minimum in the light curves was confirmed at both wavelengths and was probably due to a 
decrease in the number of UV photons available to ionize  the gas surrounding the  binary system. 
At 7 mm a very well defined peak was superimposed on the declining flux density. 
It presented maximum amplitude in 29 June 2003 and lasted for about 10 days. 
We show that its origin can be free-free emission from the  gas at the shock formed by wind-wind 
collision, which is also responsible for the observed X-ray emission. 
Even though the shock strength is strongly enhanced as the two stars in the binary system approach 
each other, during periastron passage the X-ray emission is strongly absorbed and the 7 mm 
observations represent the only direct evidence of this event. 

      \keywords{stars: individual: $\eta$ Carinae -- stars: binaries: general -- stars: winds -- radio 
continuum: general}
  }
\titlerunning{$\eta$ Carinae: mm-wave emission}
   \maketitle
%

\section{Introduction}
Even after more than a century of intensive studies, little is known about Eta Carinae as a star, 
hidden as it is by a dense cloud of gas and dust. 
Analysis of the historical optical data revealed a 5.52 year periodicity, characterized by very sharp 
dips in the light curve, lasting for only a few months \citep{damineli96}.  
Similar periodicity was found in the radial velocity of the Pa$\gamma$ and Pa$\delta$ lines, 
suggesting the existence of a binary system in a highly eccentric orbit (Damineli, Conti \& Lopes 
1997; Davidson 1997; Damineli et al. 2000), although the orbital parameters were not well defined by 
the observations. 
Damineli et al. (2000) derived an eccentricity $0.65<e<0.85$ from the radial velocity data and 
Corcoran et al. (2001a) found an even higher value, $e=0.9$, in a model where the X-ray emission is 
produced by wind-wind collisions.
The ephemerides for periastron passage was also constrained by the X-ray observations, since the 
strong periodic increase in luminosity was attributed to an increase in the shock strength as the two 
stars approached each other.
Unfortunately, the maximum in the X-ray emission is not observable, because at periastron the shocked 
region seems to be absorbed by a large H column density, which produces the pronounced dip in the 
X-ray light curve and puts constrains on the ephemerides for conjunction and opposition 
\citep{corcoran01,pittard02}. The existence of a binary system was contested by Davidson et al. 
(2000), based on high spatial resolution $HST$ spectroscopic observations. Recently Falceta-Gon\c 
calves, Jatenco-Pereira \& Abraham (2005) gave an alternative explanation to the X-ray absorption, 
which puts the periastron close to conjunction, in agreement with the $HST$ data.

Radio observations of $\eta$ Carinae, from 6 cm to 1.3 mm, show also the periodic, sharp dips in their 
light curves (Cox et al. 1995a; Duncan, White \& Lim 1997; Abraham \& Damineli 1999).
The continuum emission is attributed to thermal bremsstrahlung from an optically thick plasma, 
produced either in the $\eta$ Carinae wind \citep{cox95} or in an equatorial disk \citep{duncan03}.
Interferometric monitoring with the Australian Telescope Compact Array at 3 and 6 cm with 1" 
resolution, showed an elongated structure, reaching a maximum size of about 4" ($1.3\times 10^{17}$ cm 
at a distance of 2.3 kpc), which shrinks to less than 1" during the minimum in the light curve 
\citep{duncan97}.
This effect was interpreted by Duncan \& White (2003) as due to a decrease in the number of ionizing 
photons during the low excitation phase.

In this paper we report contemporaneous observations of the last minimum in the light curves, which 
started around May 2003, at two radio wavelengths: 1.3 mm and 7 mm, obtained respectively with  SEST, 
at La Silla, Chile and with the Itapetinga radiotelescope, in Brazil. The closely spaced observations 
provided not only the light curves with great precision, but also what we believe was the detection of 
radio emission from the colliding winds, highly enhanced during periastron passage. Since at that time 
the X-ray emission from the winds was totally absorbed, our observations represent the first detection 
of this phenomenon.

\section{Observations and Results}

To better understand the radio light curve, we made daily observations at 7 mm with the Brazilian 
Itapetinga radiotelescope and weekly observations at 1.3 mm with the ESO (European Southern 
Observatory) SEST radiotelescope at La Silla, Chile, during the periodic event predicted for June 
2003. 

   \begin{figure}
      {\includegraphics{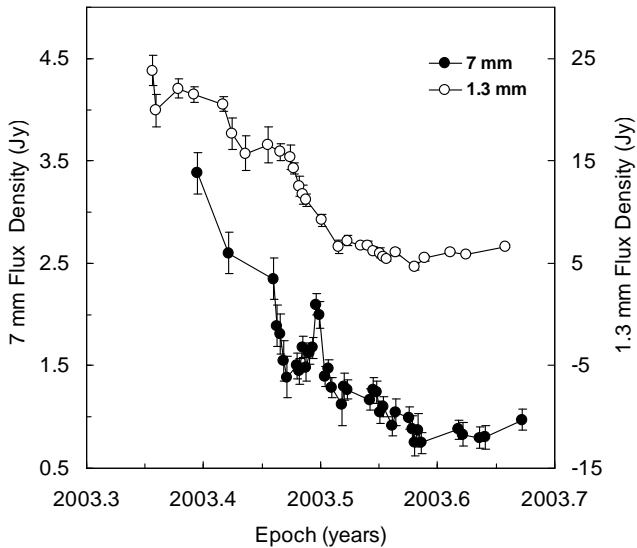}}
      \caption 
	  {mm-wave light curves of $\eta$ Carinae at the epoch of the predicted low excitation phase. Open 
circles correspond the 1.3 mm emission (right axis), full circles correspond to 7 mm emission (left 
axis).}
         \label{figure1}
   \end{figure}
   
The Itapetinga radiotelescope has a beam size of about 2 arc-min at 7 mm, enough to separate $\eta$ 
Carinae from the other HII regions in the CarII cloud; the observations consisted of scans across the 
source with  20 arc-min amplitude, to allow the subtraction of the underlying sky contribution. The 
atmospheric opacity was calculated using the technique developed by Abraham \& Kokubun (1992); the 
flux density was calibrated by observation of the standard point source Virgo A; the stability of the 
system was continuously checked by intercalating observations of the HII regions G287.57-0.59 and 
G287.55-0.63, part of the extended Carina Nebulae \citep{ret83}, with those of $\eta$ Carinae. 

Observations at 1.3 mm were made with SIMBA (SEST Imaging Bolometer Array), a 37-pixel bolometer array 
with a beamsize of $23"$ and separation between pixels in the sky of $44"$. The observations were made 
in fast scanning mode ($80"$ s$^{-1}$ in elevation). The peak flux of $\eta$ Carinae was determined by 
fitting a two-dimensional gaussian to the source; series of 4-8 maps were obtained on each occasion, 
the uncertainty in peak flux was obtained from the rms value of the fluxes in each map. The 
atmospheric opacity was determined from sky dips and daily observations of Uranus provided absolute 
calibrations. The last observation coincided with the end of the operations of the SEST 
radiotelescope.

The data, presented in Figure 1, confirms the appearance of the predicted dip, with a decrease in flux 
density by a factor of four at both wavelengths. 
The closely spaced observations revealed also a short duration peak at 7 mm, reaching the maximum flux 
density in June 29 and lasting for about 10 days, superimposed to the fast decline in emission. 
The peak is not conspicuous  at 1.3 mm, although the time sampling at this wavelength is not as dense 
as at 7 mm.
Although the general behavior of the light curves at radio frequencies is similar to what was observed 
in the previous cycle, made with coarse temporal resolution \citep{abraham99,cox99}, the short-lived 
peak observed at 7 mm was unexpected and difficult to explain in terms of, for example, a sudden 
increase in the number of ionizing photons.
We thus investigated if the excess radio emission could be produced in the thin shock region that is 
responsible for the observed X-ray emission.

\section{Model for the mm-wave emission}  
To interpret quantitatively the mm-wave light curves we will assume that the free-free emission is 
produced both in an extended disk surrounding $\eta$ Carinae \citep{duncan97} and in the cooling shock 
region formed by wind-wind collision. We will further assume that the flux of UV radiation that 
maintains the disk ionized, suddenly decreases, or maybe even stops, as proposed by Duncan \& White 
(2003).

\subsection{Emission from the optically thick disk}

If the ionizing flux decreases, part of the ionized material will recombine at a rate $r_{\rm rec}$ 
given by:

\begin{equation}
\medskip
r_{\rm rec} = \alpha ^{(1)} n_{\rm e},
\medskip
\end{equation}

\noindent
where $n_{\rm e}$ is the electron number density and $\alpha^{(1)}$ the recombination coefficient, 
given by:

\begin{equation}
\medskip
\alpha^{(1)} = \frac {2.06\times10^{-11}Z^2}{T^{1/2}}\Psi(T/Z^2) \; \; \; {\rm cm^3 \; s^{-1}},
\medskip
\end{equation}

\noindent
with the function $\Psi(T/Z^2)$ tabulated in Spitzer (1978). 
In the inner region of the disk, electron densities $n_e\sim 10^7$ cm$^{-3}$ and temperatures  $T \sim 
8000$ K are required  to explain the large flux densities and optically thick continuum radio 
spectrum, as well as the H30$\alpha$, H35$\alpha$ and H40$\alpha$ recombination lines, which present 
strong maser amplification \citep{cox95b,abraham02}. 
Using these values in equations (1) and (2) we find a recombination timescale of 2.5 days. 
Since the disk density decreases with distance to $\eta$ Carinae \citep{cox95,abraham02}, the cooling 
time will be larger in the outer parts of the disk, explaining qualitatively the duration and shape  
of the declining part of the light curve. 
The net result of this process is the formation of a hole of neutral material in the center of the 
ionized disk. 
Since the actual electron density distribution in the disk is not known and we are mainly interested 
in the residual emission, we adjusted the  7 mm flux density emitted by  the disk by a smooth function 
of the ephemerides $t$:

\begin{equation}
\medskip
S_{\rm disk}(7 \; {\rm mm}) = a\; {\rm exp}\, (b\,t+c) + m\,t+n.
\medskip
\end{equation}

The parameters $a=3.1\times 10^{-5}$, $b=-0.228$, $c=7.50$, $m= 0.004$ and $n = 0.94$, were determined 
from the data at both sides of the peak, the time $t$ was measured from 29 June 2003 $({\rm 
JD}=2452829)$.

   \begin{figure}
   \centering
   \includegraphics[bb=100 360 320 755,width=8cm,clip]{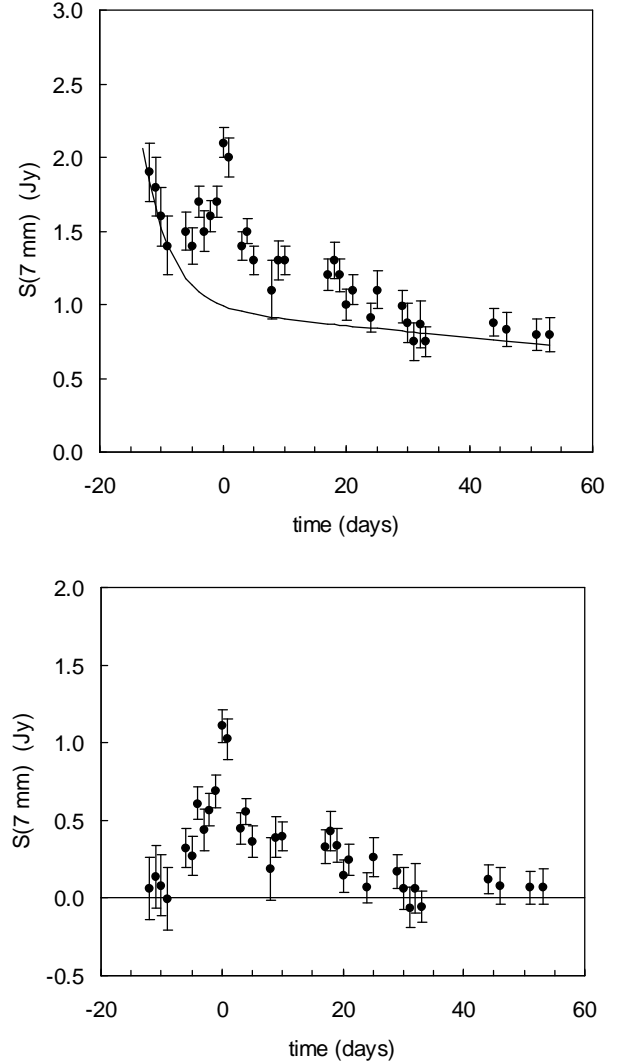}
   \caption{Expanded view of the 7 mm light curve. Upper panel: observed 7 mm flux density (filled 
circles) and model disk emission (continuous line). Lower pannel: residuals of the observed flux 
density after subtracting the disk contribution. Zero in the time axis corresponds to June 29, 2003 
(JD = 2452819).}
         \label{figure 2}
   \end{figure}
In Figure 2 we show an expanded version of the 7 mm light curve, with the assumed disk flux density 
indicated by a continuous line (upper panel). The residual emission, which origin we will try to 
establish, is seen in the lower panel.   
   
\subsection{Free-free emission from the ionized winds}

Let us assume that the winds are completely ionized and their temperatures are close to $10^4$ K; 
their densities $\rho$ at a distance $r$ from the corresponding star are related to their mass loss 
rates $\dot M$ and terminal wind velocities $v$ by:

\begin{equation}
\rho = \frac{\dot M}{4\pi r^2 v}.
\end{equation}

The flux density at mm wavelengths  must be calculated taking into account radiation transfer since, 
at the densities and temperatures found close to the stars, the optical depth can be large. Using the 
Rayleigh-Jeans approximation we have:

\begin{equation}
S(\lambda)=\frac{2kT}{\lambda^2}(1-e^{-\tau (\lambda)})\, \Omega,
\end{equation}

\noindent
where $\Omega$ is the solid angle subtended by the emitting region and $\tau (\lambda)$ the optical 
depth.

To explain the observed flux density at the peak of the emission light curve (about 1 Jy), in the most 
favorable condition of $\tau \gg 1$, it is necessary a source subtending a solid angle of 
$6\times10^{-13}$ srd, which at the distance of $\eta$ Carinae corresponds to a radius $r_0 =5.3\times 
10^{15}$ cm.

The optical depth of the emitting region is found from the integral:

\begin{equation}
\tau (\lambda)= \int^{r_0}_{r_0-\Delta r} \kappa_{ff}(\lambda)\, ds,
\end{equation}

\noindent
where  $\kappa_{ff}(\lambda)$ is the free-free absorption coefficient, which at radio wavelengths can 
be calculated from:

\begin{equation}
\kappa_{ff} (\lambda)=2\times  10^{-23}\frac{ n_i n_e g_{ff}(\lambda)\lambda^2}{T^{3/2}},
\end{equation}

\noindent
where $g_{ff}(\lambda)$ is the Gaunt factor for free-free emission. For the physical conditions in the 
winds, $T<10^5$ K and $hc/\lambda \ll  kT$, $g_{ff}(\lambda)$ can be approximated by:

\begin{equation}
g_{ff}(\lambda) = \frac{\sqrt{3}}{\pi}\biggl[17.7 +  \ln
 \biggl(2.2\frac{kT\lambda}{hc}\biggr)\biggr].
\end{equation}
\noindent
For $T>10^5$ K:

\begin{equation}
g_{ff}(\lambda) = \frac{\sqrt{3}}{\pi} \ln \biggl(2.2\frac{kT\lambda}{hc}\biggr).
\end{equation}

Taking $\Delta r_0$ as the depth of the shell that gives $\tau = 1$, we find from equations (4), (6), 
(7) and (8) $\Delta r = 3.7\times 10^{15}$ cm. 

If we assume that the peak of the 7 mm emission is due to an increase in the mass loss rate of the 
star, the perturbation will reach the radius $r = 5\times10^{15}$ cm in approximately 2 years. Since 
the observed  time scales are much smaller, we can eliminate the winds as the emission source.
We investigate now the possibility of the emission being produced by the wind-wind collision.

\subsection{Free-free emission from shocked material}

The physical conditions in the shock produced by wind collision were calculated analytically (e.g. 
Girard \& Wilson 1987; Usov 1992) and applied to the $\eta$ Carinae binary system by Ishibashi et al. 
(1999) and Corcoran et al. (2001a). Detailed  numerical simulations were obtained by Pittard et al. 
(1998) and Pittard \& Corcoran (2002), including the effects of radiative cooling. In all cases the 
spectrum of the X-ray emission and the resulting light curve  were satisfactorily explained, using 
appropriate parameters for the orbit, mass loss rates and wind velocities.
 
The radio emission from wind-wind collisions was calculate by Dougherty et al. (2003), using numerical 
models and including radiation transfer effects and different inclination angles for the orbital 
plane, to explain the thermal and non-thermal emission  detected in several binary systems (Dougherty 
et al. 1996; Dougherty \& Williams 2000; Moran et al. 1989; Churchwell et al. 1992; Williams et al. 
1997; Niemela et al. 1998). 

However, close to periastron passage and because of the very high  mass loss rate from $\eta$ Carinae 
(Hillier et al. 2001) and the high eccentricity of the binary system (Damineli et al. 2000; Pittad \& 
Corcoran 2002), radiative cooling is very fast and the shock region becomes very narrow. 
For that reason, the temperature, density and ionization profiles obtained from numerical models (e.g. 
Pittard \& Corcoran 2002), in which the cell size is about $3\times 10^{12}$ cm, cannot be used for 
our purpose. 
Therefore, we will calculate the expected radio flux density at  7 mm  with plausible  combinations of 
temperature, density and size of the emitting region.

We  first determine the physical conditions at the shock  using the expressions presented by Usov 
(1992). The momentum balance surface  intercepts the line joining the two stars  at a distance $r_1$ 
from $\eta$ Carinae, given by: 
   
\begin{equation}
\medskip
r_{1}=\frac{D}{1+\eta^{\frac{1}{2}}},
\medskip
\end{equation}
where $D$ is the distance between the  stars and $\eta=\dot{M_s} v_s/\dot{M_p} v_p$. $\dot{M_p}$ and 
$\dot{M_s}$ are the mass loss rates of $\eta$ Carinae and the companion star, $v_p$ and $v_s$ their 
respective wind velocities. Pittard \& Corcoran (2002) favored $\eta=0.2$ in their fitting of the 
X-ray spectra obtained with $Chandra$.
   \begin{figure}
      {\includegraphics{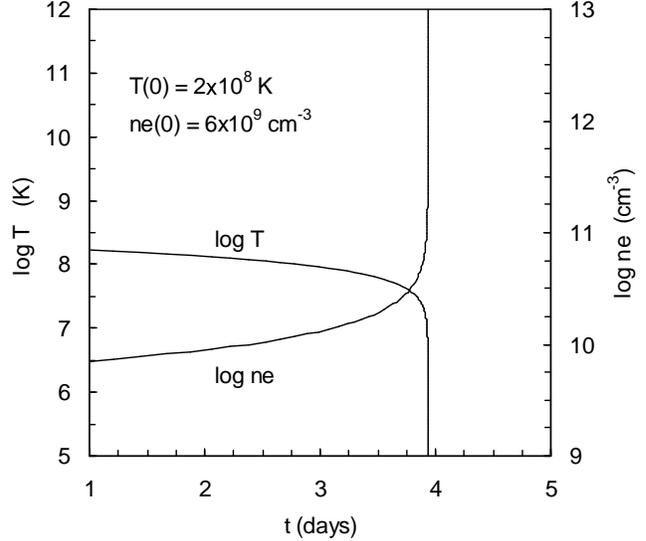}}
      \caption{Time evolution  of the temperature  and  density  behind the shock, due to energy 
losses by free-free and line emission, assuming pressure equilibrium. The initial conditions are: 
$T(0) = 2\times 10^8$ K and $n(0) = 6\times 10^{9}$ cm$^{-3}$.}
	 \end{figure}

Shocks are formed at both sides of the contact surface.
The density $\rho_2$ and temperature $T_2$ behind each shock can be derived from the values $\rho_1$ 
and $T_1$ at the shock front:

\begin{equation}
\frac {\rho_{2}}{\rho_{1}}=\frac{(\gamma +1)M^2}{(\gamma -1)M^2+2},
\end{equation}
\begin{equation}
\frac{\rho_{2}T_{2}}{\rho_{1}T_{1}}=\frac{2\gamma M^2-(\gamma -1)}{\gamma +1},
\end{equation}
where  $\gamma$ is the adiabatic exponent, $M \equiv v_{1}\left( \gamma kT_{1}/\mu m_{H} 
\right)^{-\frac{1}{2}}$ is the Mach number, $\mu$ the molecular weight and $m_H$ the mass of the H 
atom.
      
 Assuming for $\eta$ Carinae a mass loss rate of $2.5\times 10^{-4}$ M$_\odot$ yr$^{-1}$, a wind 
velocity of 700 km s$^{-1}$  and a separation $D= 1.6$ AU at periastron, corresponding to an orbit 
with eccentricity $e = 0.9$ \citep{pittard02,hillier01}, we find a number density $n_1 = \rho_1/\mu 
m_{\rm H} = 5\times 10^{10}$ cm$^{-3}$.
For a temperature $T_1= 10^4$ K,  the post-shock physical parameters will be $n_2 = 2 \times 10^{11}$ 
cm$^{-3}$ and $T_2 = 6\times 10^6$ K.
For a secondary star with mass loss rate $\dot M_s = 10^{-5}$ M$_\odot$ yr$^{-1}$ and wind velocity 
$v_s = 3000$ km s$^{-1}$, the corresponding values behind the shock are $n_2 = 6 \times 10^{9}$ 
cm$^{-3}$ and $T_2 = 2\times 10^8$ K, temperature  compatible with $Chandra$ X-ray observations 
\citep{seward01,corcoran01b}.

With these initial values, we can calculate the temperature and density time evolution  from the first 
law of thermodynamics assuming, as in Mathews \& Doane (1990), isobaric equilibrium  and  cooling by 
free-free and line emission:

\begin{equation}
\frac{3k\rho(t)}{2\mu m_{H}} \frac{dT(t)}{dt}-\frac{kT(t)}{\mu m_{H}}\frac{d\rho(t)}{dt}=-\Lambda.
\end{equation}

\noindent
with $\Lambda$ is giveny by:

\begin{equation}
\Lambda =\biggl[2.4\times 10^{-27}T^{\frac{1}{2}} + \frac{b_1 T^{p}}{(1+b_2T^{q})}\biggr]n_i n_e \; \; 
{\rm {  erg\; cm^{-3}\; s^{-1}}},
\end{equation}

\noindent
where $ n_ i$ and $n_e$ are the number density of ions and electrons, respectively, and $b_1=1.53 
\times 10^{-27}$, $b_2=1.25 \times 10^{-9}$, $p=1.2$ and $q=1.85$.
We did not consider heating by the stellar UV radiation because most of the radio emission is produced 
at temperatures larger than $10^5$ K, when this heating process is not effective.

In Figure 3 we show the time evolution of $n_e$ and $T$, for the secondary shock and the physical 
conditions in the post-shock gas derived previously. We can see that the temperature decreases slowly 
during approximately 4 days and then drops very fast, as the density increases and line emission 
becomes effective. For the primary shock, the cooling time is less than a day, since the initial 
temperature is lower and the density much higher.

As in the case of the individual wind emission, we can estimate the size of the  optically thick  
region necessary to produce the observed peak at 7 mm. Figure 4 shows its radius as a function of the 
electron temperature, for the conditions found in the secondary shock. This size will be 4 times 
larger for the primary shock. We should  notice that $\tau$ cannot be much larger than one at 7 mm, 
otherwise the optical depth at 1.3 mm could also be large, in which case the expected flux density 
would be 28 Jy, much larger than what was observed. In the same Figure we show for this reason, the 
depth $L$ of the emitting region  that gives $\tau = 1$ both for 7 mm and 1.3 mm. The allowed values 
of $L$ lie between the two lines. The dotted line represents the separation between the secondary star 
and the shock at periastron, which gives an upper limit to the depth and temperature of the material 
responsible for the peak emission. The same calculation can be made for the primary shock, although in 
this case, for the same emitting area the emission  is 16 times smaller.

   \begin{figure}
      {\includegraphics{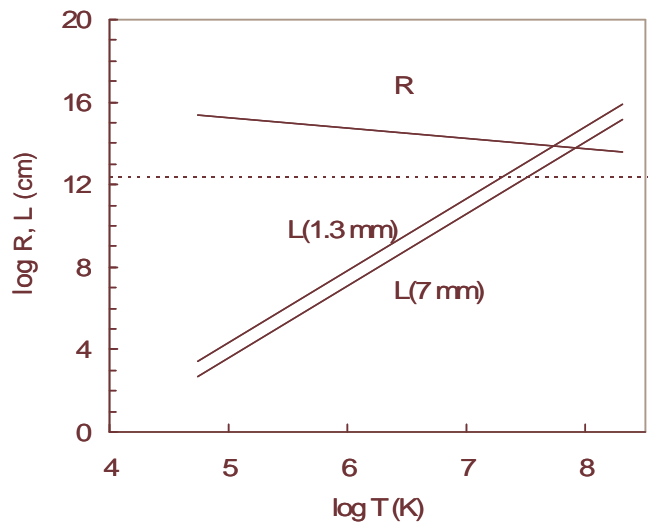}}
      \caption 
	  {Radius $R$ and depth $L$ of the emitting region that reproduces the peak flux density and $\tau 
= 1$, respectively, as a function of the electron temperature. The dotted line represents the 
separation between shock and star.}
         \label{figure4}
   \end{figure}

\section{Conclusions}

In this paper we presented the 1.3 and 7 mm light curves of $\eta$ Carinae during the 2003.5  low 
excitation phase. The expected minimum  in the light curves was confirmed but an unexpected peak, 
superimposed to the decreasing flux density, was observed at 7 mm. It reached its maximum in 29 June 
2003 and lasted for about 10 days. To interpret the light curves we assumed that the decreasing part 
of the emission is produced in the extended disk surrounding the $\eta$ Carinae binary system. We 
tested the possibility of the peak emission being produced either by the single stars winds or by the 
wind-wind shock region, also responsible for the X-ray emission.
We calculated the size of the emitting region assuming single wind emission and obtained a radius of 
$5\times 10^{15}$ cm and a depth about $3.7\times 10^{15}$ cm for $\tau =1$. Considering the wind 
velocity of 700 km s$^{-1}$,  a change in mass loss rate would take about 20 years to reach this 
distance.
An alternative was to investigate the emission from the wind-wind collision. We  calculated  the 
physical conditions in the material behind the primary and secondary shocks and estimated the cooling 
time due to free-free and line emission. The size of the emitting  region was determined as a function 
of temperature, as well as the possible  depth range compatible with both  7 and 1.3 mm observations.

\section*{Acknowledgments}

      This work was supported by the Brazilian Agencies FAPESP, CNPq and FINEP.

\end{document}